\documentclass[12pt]{article}
\usepackage{graphicx,multirow}
\usepackage{latexsym}
\usepackage{amsmath,amsfonts,amssymb}
\begin{document}
\begin{center}
{\bf FINITE TEMPERATURE CASIMIR EFFECT AND DISPERSION IN THE
PRESENCE OF COMPACTIFIED EXTRA DIMENSIONS }

\vspace{1cm} Marianne
Rypest{\o}l\footnote{marianne.rypestol@fys.uio.no}  and Iver
Brevik\footnote{iver.h.brevik@ntnu.no}

\bigskip
Department of Energy and Process Engineering, Norwegian University
of Science and Technology, N-7491 Trondheim, Norway

\vspace{1cm}

\end{center}

\begin{abstract}
Finite temperature Casimir theory of the Dirichlet scalar field is
developed, assuming that there is a conventional Casimir setup in
physical space  with two infinitely large plates separated by a
gap $R$ and in addition an arbitrary number $q$ of extra
compacified dimensions. As a generalization of earlier theory, we
 assume in the first part of the paper that there is a scalar 'refractive index' $N$ filling
the whole of the physical space region.  After presenting general
expressions for free energy and Casimir forces we focus on the low
temperature case, as this is of main physical interest both for
force measurements and also for issues related to entropy and the
Nernst theorem. Thereafter, in the second part we analyze dispersive properties,
 assuming for simplicity $q=1$,  by taking into account dispersion associated with
the first Matsubara frequency only. The medium-induced contribution to the free energy, and pressure, is calculated at low temperatures.

\end{abstract}

PACS numbers: 11.10.Kk, 11.10.Wx, 42.50.Lc

\section{Introduction}
Consider two infinite parallel plates separated by a gap $R$, and
assume that the  field between the plates, as well as  on the
outside, is a scalar field obeying Dirichlet  boundary conditions
at $z=0$ and $z=R$ ($z$ is the direction normal to the plates). In
order to get a Casimir configuration  we have to take into account
  the field outside the plates also. Let one of the plates,
the right one at $z=R$, say, be denoted a "piston". Now generalize
the situation such that $p$ spatial dimensions are envisaged,
together with $q$ extra compactified dimensions. The spacetime
dimension is thus $D=p+q+1$. We are led to a Casimir piston model
in which spacetime is {\it flat}. This model has attracted
considerable attention in the recent literature
\cite{elizalde09,teo09c,lim08,lim08a,teo09,teo09a,teo09b,fulling09,kirsten09,cheng08,
cheng06,cheng06a,cheng05,cheng05a,edery08,edery08a,marachevsky07,frank08,fulling08a,poppenhaeger04,cavalcanti04,teoI,teoII,teoIII}.
One reason for the current interest is obviously  the mathematical
elegance of the formalism; the efficiency of regularization
procedures like the zeta function regularization is quite
striking. Typical for this kind of theories is that the field
energy can be expressed in terms of Epstein-like zeta functions.
The Casimir force between the two plates in physical space follows
by taking the derivative of the energy (or free energy at finite
temperature) with respect to $R$.

Another motivation is of a more physical nature, namely to
investigate constraints for non-Newtonian gravity from the Casimir
effect (cf., for instance, Refs.~\cite{bordag01} and
\cite{milton04}). Present Casimir force experiments are so
accurate that the possible influence from extra dimensions is
taken seriously. The hypothetical extra force is usually taken to
be of the Yukawa form.

In the present paper we will however not consider possible
Casimir-induced deviations from Newtonian mechanics, but focus
instead on the analysis of Casimir free energy and force in the
presence of extra compactified dimensions at {\it finite}
temperature  $T$. At the final stages of our work, we actually
became aware of a series of papers of Teo, which go along similar
lines as our considerations do. There is necessarily some overlap
between Teo's papers and the present one.  However, since our
results were for the most part obtained  independently of each
other, it may be of interest to make a more close comparison:

 In Ref.~\cite{teo09}, the finite temperature Casimir force in
the presence of extra dimensions was calculated for the scalar
field with Dirichlet boundary conditions. The free energy was not
considered. The force was always found to be attractive. When the
ratio $R/L$ (in our notation) increased from 0.1 to 1 the Casimir
force was found to increase quite much, by a factor of about 3. In
our approach, the Casimir force is calculated via the free energy.
A satisfactory feature is that a detailed comparison (some
calculation is needed) shows  our results to be in agreement with
each other.

  A more extended version of this work was given in the next
paper  \cite{teo09a}, considering also the free energy. In the
third paper in the series \cite{teo09b}, the free energy was
studied for a massive scalar field with general curvature coupling
subject to Dirichlet or Neumann boundary conditions, thus a more
general setting than that considered in our present investigation.

Finally, in three recent papers Teo investigated the finite
temperature effect for a scalar field in multidimensional space
assuming {\it Robin} boundary conditions \cite{teoI}; then gave a
comparison between Kaluza-Klein and Randall-Sundrum models with
special emphasis on the non-reversal of sign in the presence of
warped extra dimensions \cite{teoII}, and worked out the Casimir
theory for a multi-layer dielectric slab model in piston geometry
\cite{teoIII}.

\bigskip

In the present paper we focus on the following two issues:

\bigskip

1) We allow there to be a 'refractive' index, called  $N$,
experienced by the scalar field in the physical medium between the
plates. This influences the Casimir free energy as well as  the
entropy, and is of interest in connection with the Nernst theorem
as $T\rightarrow 0$. To our knowledge, such a generalization has
not been considered before. The introduction of a refractive index
is not so trivial as one might imagine, if one compares with
electrodynamic theory: in that case, the Casimir free energy
expression $F$ gets an extra factor $N$ in the denominator at
$T=0$, the finite $T$ corrections being more complicated. Here the
$T=0$ behavior can be understood in a physical way by observing
that the momentum of a photon in the medium can be written in the
form $\sqrt{k_\perp^2+\pi^2m^2/R^2}$, where ${\bf k}_\perp$ is the
photon momentum parallel to the plates whose separation is $R$,
and $m$ is an integer (in turn, this is intimately linked to the
assumption that the photon momentum is as given by the Minkowski
energy-momentum tensor). Taking into account the volume element in
phase space, one obtains the essential result for $F$. Cf.
Ref.~\cite{brevik08}, and also the remarks of Ravndal
\cite{ravndal08,ravndal09}.

2)  Our second point is to take into account the frequency
dispersion of the medium in a crude way, by including the
refractive index for low frequencies only. Specifically, we cut
off the influence from the medium, thus setting $N=1$, for all
frequencies exceeding the first Matsubara frequency $\zeta_1=2\pi
T$. From a basic viewpoint this procedure is permitted, since
there is no conflict with the fact that the refractive index -
like any other generalized susceptibility -  has to be such that
it decreases monotonically along the positive imaginary frequency
axis. Moreover, in physical units $\zeta_1$ is lying in the region
$10^{14}~\rm{s}^{-1}$ and is thus quite high, making the model
sensible physically. This generalization is made for the case
$q=1$ only.

We begin in the next section by considering the general case where
the number $p$ of edges in physical space is allowed to take  the
values 1, 2, or 3. Thereafter we specialize to the case $p=1$,
corresponding to the conventional setup with two parallel plates
separated by a gap $R$. Zeta function regularization is employed.
 Our main focus will be on low temperatures as mentioned, as this
is of main physical interest. For $N$ constant we evaluate the
Casimir free energy, the corresponding  pressure, the entropy, and
work out some simplifying limiting expressions. The dispersion
generalization is considered in section 6.

Finally we mention for completeness that  all plates, in physical
space as well as in the extra compactified space, are so large
that edge effects are negligible. We use natural units, with
$\hbar=c=k_B=1$.

\section{General formalism, when $p=1,2,3$}

We  assume finite temperatures  from the outset. Let
  $p$ be the number of edges in
physical space. Thus $p$ can take the values  $p=1$ (two plates),
$p=2$ (four plates), or $p=3$ (box). The number $M$ of transverse
dimensions in physical space  is  $M=3-p$. We let index $i_\perp$ refer
to the transverse directions, so that $i_\perp \in [1,M]$. The number
$q$ of extra compactified dimensions is at first assumed
arbitrary. The total spacetime dimension $D$ is thus $D=4+q$. We
assume there to be a scalar field $\Phi(x^i, y^j, t)$ in the bulk.
As mentioned, Dirichlet boundary conditions are assumed on the
plates in physical space. The plate separations are $R_i$, with
$i\in [1,p]$. In the  extra space with dimension $q$ we assume, in
accordance with common usage, that a torus of circumference $2\pi
L_j$, $j \in [1,q]$, is attached to each spacetime point. Periodic
boundary conditions are assumed for the extra dimensions.

As mentioned, we will allow for a constant 'refractive index' $N$
in the physical bulk. This index  describes the response from the
 homogeneous and isotropic medium  to the scalar
field, in  the same way as  the usual refractive index
$\sqrt{\varepsilon \mu}$ does in a dielectric medium,
$\varepsilon$ being the permittivity and $\mu$ the permeability.
Our model is thus a sort of scalar electrodynamics which has been
considered occasionally in earlier studies  also, for instance in
Ref.~\cite{brevik87}, dealing with curvilinear space.  In the
ordinary case $D=3$ our model implies that the scalar photons
possess the momentum ${\bf k}=N\omega \hat{\bf k}$, in accordance
with the Minkowski momentum in dielectric media in ordinary
electrodynamics as already mentioned, implying  a spacelike total
photon four-momentum.

The central differential operator in Euclidean space $(\tau=it)$
is $\Box_E=N^2\partial_\tau^2+\sum_i\partial_i^2+\sum_j
\partial_j^2$. The  free energy $F$ can be found by using the
zeta function for the operator $-\Box_E$,
\begin{equation}
\zeta_{-\Box_E}(s)=\sum_J\lambda_J^{-s}, \label{3}
\end{equation}
where in the present generalized case  the eigenvalues are
\begin{equation}
\lambda_J=k_\perp^2+\sum_{i=1}^p\left(
\frac{n_i\pi}{R_i}\right)^2+\sum_{j=1}^q\left(\frac{m_j}{L_j}\right)^2+N^2\zeta_l^2.
\label{4}
\end{equation}
Here  $J$ is an index referring to all the indices $\{n_i\},
\{m_j\},\{k_t\}$, as well as the Matsubara frequencies $\zeta_l=2\pi lT$ with
$l=0,1,2,...$. The Dirichlet
conditions on the plates causes the first sum to run over positive
$n_i$ only. Further, $m_j$ extends over all integers because of
periodic boundary conditions. In (\ref{4}) we have also introduced
$k_\perp^2$ as the square of the transverse wave number ${\bf
k}_\perp$, i.e., $k_\perp^2=\sum_{i_\perp=1}^M k_{i_\perp}^2.$ We emphasize that
$N$ refers to the physical space only. To introduce an analogous
refractive index in the extra space would hardly have any physical
meaning.

The general expression for the free energy reads
\begin{equation}
F=-\frac{T}{2}\zeta'_{-\Box_E}(0). \label{6}
\end{equation}
We first consider the interior free energy, called $F_I$, in the
cavity. With $V_\perp$ denoting the transverse volume,
$V_\perp=\prod_{i_\perp=1}^M R_{i_\perp}$, and with the quantity $A$ defined as
\begin{equation}
A=(2\pi N lT)^2+ \sum_{i=1}^p\left(
\frac{n_i\pi}{R_i}\right)^2+\sum_{j=1}^q\left(\frac{m_j}{L_j}\right)^2,
\label{8}
\end{equation}
we have for the zeta function
\begin{equation}
\zeta_{-\Box_E}(s)=\frac{V_\perp}{(2\pi)^M}\sum_{l=-\infty}^\infty
\sum_{\{n_i\}=1}^\infty \sum_{\{m_j\}=-\infty}^\infty \int d^M
k_\perp(k_\perp^2+A)^{-s}. \label{9}
\end{equation}
Here the notation $\{n_j\}=1 $ means that $n_1=1,2,...\infty,\,
n_2=1,2,...\infty$.
 The integral can be evaluated using the
technique of generalized polar coordinate transformations
\cite{li97}. This yields, after integrating over all angles,
\begin{equation}
\zeta_{-\Box_E}(s)=\frac{V_\perp}{(2\pi)^M} \frac{2 \pi^\frac{M}{2} }{\Gamma(M/2)}
\sum_{l=-\infty}^\infty \sum_{\{n_i\}=1}^\infty \sum_{\{m_j\}=-\infty}^\infty
\int_0^\infty dr ~r^{M-1}(r^2+A)^{-s}.
\label{11}
\end{equation}
After the variable change $x=r^2/A$ the integral part can be
recognized as the integral representation of the beta function
$B(v,u) = \Gamma(v) \Gamma(u)/\Gamma(v+u)$ and hence the zeta function is
\begin{equation}
\zeta_{-\Box_E}(s)=\frac{V_\perp}{(2\pi)^M} \frac{\pi^\frac{M}{2} \Gamma(s-M/2)}{\Gamma(s)}
\sum_{l=-\infty}^\infty \sum_{\{n_i\}=1}^\infty \sum_{\{m_j\}=-\infty}^\infty A^{M/2-s} .
\label{11a}
\end{equation}
The free energy can easily be found from the above expression
if we remember that $(g(z)/\Gamma(z))'_{z=0}=g(0)$ for any function $g(z)$.
\begin{figure}[ht!]
    \centering
    \includegraphics[width=\linewidth]{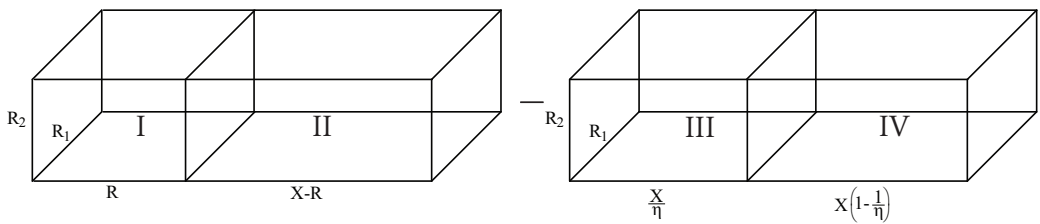}
    \caption{Illustration of the four cavities of the piston model.}
\end{figure}

We now introduce the piston in the $i=p$ direction, observing that
the Casimir free energy of the piston geometry, as usual in Casimir calculations,  consists of four
parts: $F_\mathrm{I}= F_\mathrm{I}(R_1, \ldots , R_{p-1}, R_p,
L_1, \ldots L_q)$ is the free energy of cavity I with length $R_p$
in the piston direction. For cavity II we have $F_\mathrm{II}=
F_\mathrm{II}(R_1, \ldots , R_{p-1},X-R_p, L_1, \ldots L_q)$ and
similar expression for cavity III and IV with lengths $X/\eta$ and
$X(1-1/\eta)$ in the piston direction respectively. The parameter  $\eta$ is taken
to have a  value of around  2 (see Fig.~1). The main point here is that cavity I is a squeezed region,
in which the discretization of allowable wave numbers is essential. Cavity II, on the other hand, is non-squeezed,
 so that  summations over wave numbers can be replaced by integrals. Similarly, the reference configuration (cavities
 III and IV) corresponding to $\eta \sim 2$, is also non-squeezed. The technique is explained in more detail in Ref.~\cite{elizalde94}, Ch. 4, for instance. The conventional
situation with  parallel plates is retrieved by taking $X\to \infty$.

Thus in $F_\mathrm{II}$ the sum over the discrete $k_p = n_p\pi/R_p$ goes to an integral in the limit $X\to \infty$.
This integral can be evaluated together with the integrals over the transverse directions in Eq.~\eqref{9}.
To find the free energy of cavity II we need only to replace $M$ with $ M+1$,
$ V_\perp$ with $(X-R)V_\perp $, remove $(n_p\pi/R_p)^2$ in $A$ and remove the sum over $n_p$.

 Assume hereafter that   the
compactified extra dimensions have all the same size, ($L_i=L$).
We introduce the Epstein-like zeta function
\begin{equation}
E_V(a_1,a_2,...,a_V;s) = \sum_{\{n_j\}=1}^\infty \left(
\sum_{j=1}^V a_j n_j^2 \right)^{-s}. \label{21}
\end{equation}

After some algebra we then obtain the following expression for the
 free energy, called $F$, of the piston system
\begin{equation}
\begin{split}
F =& F_\mathrm{I} + F_\mathrm{II}-F_\mathrm{III}-F_\mathrm{IV}
= -\frac{1}{2} T\pi^{M/2} \frac{V_\perp}{(2\pi)^{M}} \sum_{k=0}^q \binom{q}{k} (2)^{q-k}  \\
& \times \Bigg[ \Gamma \left(-\frac{M}{2}\right) \Big\{ \ E_{p+q-k}\left( \frac{\pi^2}{R_1^2}, \ldots , \frac{\pi^2}{R_p^2}, \frac{1}{L^2}, \ldots, \frac{1}{L^2}; -\frac{M}{2} \right) \\
& ~+ 2 E_{1+p+q-k} \left( (2\pi N T)^2, \frac{\pi^2}{R_1^2}, \ldots , \frac{\pi^2}{R_p^2}, \frac{1}{L^2}, \ldots, \frac{1}{L^2}; -\frac{M}{2} \right) \Big\} \\
-& \frac{R_p}{2\sqrt{\pi}} \Gamma \left(-\frac{M+1}{2}\right) \Big\{ E_{p-1+q-k}\left( \frac{\pi^2}{R_1^2}, \ldots , \frac{\pi^2}{R_{p-1}^2}, \frac{1}{L^2}, \ldots, \frac{1}{L^2}; -\frac{M+1}{2} \right) \\
& ~+ 2E_{p+q-k} \left( (2\pi NT)^2, \frac{\pi^2}{R_1^2}, \ldots ,
\frac{\pi^2}{R_{p-1}^2}, \frac{1}{L^2}, \ldots, \frac{1}{L^2};
-\frac{M+1}{2} \right) \Big\} \Bigg]. \label{22}
\end{split}
\end{equation}
This expression holds for arbitrary temperature, for arbitrary $p
\in [1,3]$ and  for arbitrary integers $q$, when all the $L_i$ are
equal. As is usual,  $F$ refers to unit surface area. Recall that
the dimensions of the plates in the transverse dimensions are
assumed infinite.

\section{The case $p=1$}

When $p=1$, the system in physical space consists of  two parallel
plates separated by a gap $R_p = R$. Thus $M=2$. The value of $q$
is at present kept arbitrary. With $\Gamma(-3/2)=(4/3)\sqrt{\pi}$
we get
\begin{equation}
\begin{split}
F =& - \frac{T}{8\pi} \sum_{k=0}^q \binom{q}{k} (2)^{q-k} \Bigg[\Gamma \left(-1\right) \Big\{ \ E_{1+q-k}\left( \frac{\pi^2}{R^2}, \frac{1}{L^2}, \ldots, \frac{1}{L^2}; -1 \right) \\
& \quad + 2 E_{2+q-k} \left( (2\pi NT)^2, \frac{\pi^2}{R^2}, \frac{1}{L^2}, \ldots, \frac{1}{L^2}; -1 \right) \Big\} \\
& \quad - \frac{2R}{3} \Big\{ E_{q-k}\left( \frac{1}{L^2}, \ldots, \frac{1}{L^2}; -\frac{3}{2} \right) \\
& \quad + 2E_{1+q-k} \left( (2\pi NT)^2, \frac{1}{L^2}, \ldots,
\frac{1}{L^2}; -\frac{3}{2} \right) \Big\} \Bigg]. \label{23}
\end{split}
\end{equation}
The terms with $k=q$ are independent of $L$ and are equal to the
Casimir free energy in ordinary 3+1 spacetime ($q=0$). Naming
these terms $F_{q=0}$ we evaluate them separately
\begin{equation}
\begin{split}
F_{q=0} =& -  \frac{T}{8\pi} \Bigg[\Gamma \left(-1\right) \left\{
E_1\left( \frac{\pi^2}{R^2}; -1 \right)
+ 2  E_2 \left( (2\pi NT)^2, \frac{\pi^2}{R^2}; -1 \right)  \right\} \\
& ~-\frac{4R}{3} E_1 \left( (2\pi NT)^2; -\frac{3}{2} \right)
\Bigg].
\end{split}
\label{24}
\end{equation}
Note that the term $E_{q-k}\left( \frac{1}{L^2}, \ldots,
\frac{1}{L^2}; -\frac{3}{2} \right)$ in Eq.~(\ref{23}) vanishes
when $k=q$. Equation (\ref{24}) can be simplified by using
$E_1(a_1;s) = a_1^{-s} \zeta_R (2s)$, where $\zeta_R$ is the
Riemann zeta function, and the reflection formula
\begin{equation}
\Gamma \left( \frac{z}{2} \right) \zeta_R(z) = \pi^{z-1/2} \Gamma
\left( \frac{1-z}{2} \right) \zeta_R (1-z). \label{26}
\end{equation}
The Casimir free energy per surface area is then
\begin{equation}
\begin{split}
F &= F_{q=0} - \frac{T}{8\pi} \sum_{k=0}^{q-1} \binom{q}{k} (2)^{q-k} \Bigg[\Gamma \left(-1\right) \Big\{ \ E_{1+q-k}\left( \frac{\pi^2}{R^2}, \frac{1}{L^2}, \ldots, \frac{1}{L^2}; -1 \right) \\
& \quad + 2 E_{2+q-k} \left( (2\pi NT)^2, \frac{\pi^2}{R^2}, \frac{1}{L^2}, \ldots, \frac{1}{L^2}; -1 \right) \Big\} \\
& \quad - \frac{2R}{3} \Big\{ E_{q-k}\left( \frac{1}{L^2}, \ldots, \frac{1}{L^2}; -\frac{3}{2} \right) \\
& \quad + 2E_{1+q-k} \left( (2\pi NT)^2, \frac{1}{L^2}, \ldots,
\frac{1}{L^2}; -\frac{3}{2} \right) \Big\} \Bigg], \label{27}
\end{split}
\end{equation}
with
\begin{equation}
F_{q=0} = N^3T^4 \frac{R \pi^2 }{90}-T\frac{\zeta_R(3)}{16 \pi
R^2} - T\frac{\Gamma (-1)}{4 \pi} E_2 \left( (2\pi NT)^2,
\frac{\pi^2}{R^2}; -1 \right). \label{28}
\end{equation}
The Epstein-like functions in  Eqs.~(\ref{27}) and (\ref{28}) will
be zeta-function regularized  \cite{elizalde94,elizalde95}.  By
repeatedly using
\begin{equation}
\begin{split}
&E_V(a_1,a_2,\ldots,a_V;s) = -\frac{1}{2} E_{V-1}(a_2,a_3,\ldots ,a_V;s) \\
&\quad + \frac{1}{2} \sqrt{\frac{\pi}{a_1}} \frac{\Gamma \left(s-\frac{1}{2} \right)}{\Gamma (s)} E_{V-1}(a_2,a_3,\ldots,a_V;s-\frac{1}{2}) \\
& \quad + \frac{2 \pi^s}{\Gamma(s)} a_1^{-(s+1/2)/2} \sum_{n_1,n_2,\ldots,n_V=1}^\infty n_1^{s-1/2} \left( \sum_{i=2}^V a_i n_i^2 \right)^{-(s-1/2)/2} \\
& \qquad \times K_{s-1/2}\left( \frac{2\pi}{\sqrt{a_1}} n_1 \left( \sum_{i=2}^V a_i n_i^2 \right)^{1/2} \right),
\label{29}
\end{split}
\end{equation}
we can express $F$ in terms of the modified Bessel functions of
the second kind, $K_\nu$. We will now discuss limits of physical
interest.

\section{Approximations}

As the above formalism is complicated, it is of interest to make
mathematical simplifications applicable to situations of physical
interest.

There is one simplification that is  obvious, resulting from the
fact that the separations in physical space space are very much
greater than those in the extra space. Casimir measurements are
usually done with separations of order $1~\mu$m or less. In the
extra space, one can estimate the typical distances by looking for
minima of the free energy. For instance, by making for a moment a digression into warped space,
 in Ref.~\cite{brevik01}
we found that at low temperatures there is a minimum of the free
energy in a Randall-Sundrum  model if the length $r_c$ of the
orbifold is of order $10^{-18}~ \rm{GeV}^{-1}\sim 10^{-19}$ fm. It appears reasonable that we can set
\begin{equation}
R/L \gg 1
\end{equation}
for all practical purposes.

\subsection{Low temperatures}

The most important limiting case
is when  $T$ is low. This case gradually catches up the quantum
mechanical zero-point energy as $T\rightarrow 0$. We shall take
the low-temperature limit to mean
\begin{equation}
RT \ll 1. \label{30}
\end{equation}
Obviously, it implies that also $LT \ll 1$. In dimensional terms,
Eq.~(\ref{30}) implies $(k_BT)(R/\hbar c) \ll 1$, or $RT \ll
0.002$ with $T$ in kelvin and $R$ in meters. Thus to one percent
accuracy, $ R=100$ nm corresponds to $ T < 200$ K, whereas
$R=1000$ nm corresponds to $T < 20$ K.

We now use Eq.~(\ref{29}) to find the finite expression for
$F_{q=0}$,
\begin{equation}
\begin{split}
F_{q=0} =& N^3T^4 \frac{R \pi^2 }{90}-T\frac{\zeta_R(3)}{16 R^2}
+\frac{T}{8\pi} \Gamma(-1) E_1\left( \frac{\pi^2}{R^2}; -1 \right) \\
&-\frac{1}{12\pi N} E_1\left( \frac{\pi^2}{R^2}; -\frac{3}{2} \right) \\
&-\frac{1}{\sqrt{2}} \left( \frac{NT}{R} \right)^\frac{3}{2} \sum_{n,l=1}^\infty \left( \frac{n}{l} \right)^{\frac{3}{2}} K_{\frac{3}{2}}
\left( \frac{\pi l n}{NTR} \right) \\
=& -\frac{\pi^2}{1440NR^3} + N^3T^4 \frac{R \pi^2
}{90}-\frac{1}{\sqrt{2}} T \sqrt{ \frac{NT}{R^3} }
 \sum_{n,l=1}^\infty \left( \frac{n}{l} \right)^{\frac{3}{2}} K_{\frac{3}{2}} \left( \frac{\pi l n}{NTR} \right)
\end{split} \label{31}
\end{equation}
Notice that $\frac{T}{8\pi} \Gamma(-1) E_1\left(
\frac{\pi^2}{R^2}; -1 \right)$ cancels $-T\frac{\zeta_R(3)}{16 \pi
R^2}$, coming from the Matsubara integer equal to zero ($l=0$), in
view of  Eq.~(\ref{26}). This equation does so far not imply the
low temperature limit. But since  the argument in $K_\nu$ is
large, we need only the first term of the asymptotic expansion
$K_{\nu}(z)= (\pi/2z)^{1/2}e^{-z}$. Due to the product $nl$ in the
exponent of the expansion we only need the dominant term $n=l=1$
in the double sum. Thus $F_{q=0}$ is
\begin{equation}
F_{q=0} = - \frac{\pi^2}{1440N R^3} + N^3T^4 \frac{R \pi^2 }{90} -
NT^2 \frac{e^{-\frac{\pi}{NRT}} }{2R} \label{33}
\end{equation}
in the low temperature limit. Here $T^4 \frac{R \pi^2 }{90}$ is
the Stefan-Boltzmann term corresponding to the vacuum energy in
empty 3+1 spacetime and $-\frac{(NT)^2}{2 R}e^{-\frac{\pi}{NRT}}$
is the leading correction term. One may notice that there is no
term proportional to $T^3 \zeta_R(3)$; this term being
characteristic for  electromagnetic fields.

As regards the dependence upon $N$, the following points may be
noted. At $T=0$ the (free) energy is one half the sum over
eigenfrequencies $\omega$. In view of the Dirichlet boundary
conditions on the plates this sum becomes transferable to a sum
over integers, which may conveniently be done in view of zeta
function regularization. As $\omega=k/N$, the proportionality
factor $1/N$ in the first term in Eq.~(\ref{33}) necessarily
follows. The second term in the same equation involves  the
thermal occupation number $1/(e^{\omega /T}-1)$. This does not
contain $N$ in itself, but the refractive index occurs again in
the phase space element $d^3k/(2\pi)^3$, which brings in a factor
$N^3$. This is seen to agree with Eq.~(\ref{33}). The other terms
in Eq.~(\ref{31}) are more complicated, but it is useful to
observe that $F$ is generally for finite $T$ equal to $T$
multiplied with the logarithm of the partition function. As $N$
occurs only in the combination $(NT)$ in the quantity $A$ defined
in Eq.~(\ref{8}), which in turn is closely related to the
eigenvalues of the basic $\lambda_J$ operator in Eq.~(\ref{4}),
the ratio $F/T$ has to be a function of  $(NT)$. We see that this
is the case for all terms in Eqs.~(\ref{31}) and (\ref{33}). Cf.
also the remarks of Ravndal \cite{ravndal08,ravndal09}.

To find the low temperature limit of the part of the free energy
depending on $L$ and $q$ it is convenient to  evaluate the
remaining part of Eq.~(\ref{27}) in pairs. In this way we obtain
after a lengthy algebra (details omitted) the following expression
for the free energy in the low temperature limit

\begin{equation}
\begin{split}
F =& - \frac{\pi^2}{1440 NR^3} + \sum_{k=0}^{q-1} \binom{q}{k} (2)^{q-k}\Bigg[ \frac{R}{32 \pi^2 N} \Gamma (-2) E_{q-k} \left(\frac{1}{L^2},\ldots,\frac{1}{L^2};-2\right) \\
& \quad - \frac{1}{12 \pi N} E_{1+q-k}\left( \frac{\pi^2}{R^2},\frac{1}{L^2},\ldots,\frac{1}{L^2};-\frac{3}{2} \right) \Bigg]\\
&+N^3T^4 \frac{R \pi^2 }{90} - NT^2 \frac{e^{-\frac{\pi}{NRT}} }{2 R}+ 2qRT \left( \frac{NT}{2\pi L}\right)^{\frac{3}{2}} e^{-\frac{1}{NTL}} \\
&-  q \frac{NT^2}{\pi}  \sqrt{\frac{\pi^2}{R^2} + \frac{1}{L^2}}
\exp\left(-\frac{1}{NT} \sqrt{\frac{\pi^2}{R^2} + \frac{1}{L^2}}
\right). \label{36}
\end{split}
\end{equation}
This is our main result. At zero temperature only the first three
terms remain. The result agrees with Ref.~\cite{cheng06} when
$N=1$, except from terms in that reference leading to repulsive
forces. We are, as mentioned, concerned  with the piston model,
where the repulsive terms are removed. This is also the situation
considered in Ref.~\cite{cheng08}.

Again using Eq.~(\ref{29}) we can rewrite the zero-temperature
energy $E$  per unit area as
\begin{equation}
\begin{split}
E &= - \frac{\pi^2}{1440 NR^3} + \sum_{k=0}^{q-1} \binom{q}{k}
(2)^{q-k}  \\
&\times \Bigg\{\frac{1}{24\pi N} E_{q-k}(\frac{1}{L^2},\ldots,\frac{1}{L^2};-\frac{3}{2})\\
& \quad - \frac{1}{8 L^2 R \pi^2 N}
\sum_{m_1,m_2\ldots,m_{q-k},n=1}^\infty n^{-2} \left(
\sum_{j=1}^{q-k} m_j^2 \right)
 K_2\left(2 n \frac{R}{L} \left( \sum_{j=1}^{q-k} m_j^2 \right)^{1/2} \right) \Bigg\} \label{37}
\end{split}
\end{equation}
(at $T=0$ the thermodynamic energy $E$ is the same as the free
energy $F$). The Epstein-like terms are independent of $R$ and
will not influence the Casimir force. The arguments of the
modified Bessel functions are proportional to $R/L \gg 1$ and will
only give exponentially small corrections to the first term.

The two last terms in Eq.~(\ref{36}) are  low temperature
corrections from the compactified extra dimensions. They are
proportional to the number
  $q$ and are exponentially decreasing in $1/T$.

The Casimir force $P$ per unit area can be found as
 $P=-\partial F/\partial R$.
  Since $ K'_\nu (z) = -\frac{1}{2} [ K_{\nu-1} (z)
+ K_{\nu+1} (z)]$  we find, using Eq.~(\ref{31}),
\begin{equation}
\begin{split}
P_{q=0}=&-\frac{\partial  F_{q=0}}{\partial R} = - \frac{\pi^2
}{480 NR^4} -  N^3T^4 \frac{\pi^2}{90} - \frac{1}{\sqrt{2}}
 T \sqrt{ \frac{NT}{R^3} }  \\
& \times \sum_{n,l=1}^\infty\Bigg[ \frac{3n^\frac{3}{2}}{2R l^\frac{3}{2}} K_\frac{3}{2} \left( \frac{\pi l n}{NTR} \right) -
\frac{\pi n^\frac{5}{2} }{2 NT l^\frac{1}{2} R^2} \left( K_\frac{1}{2} \left( \frac{\pi l n}{NTR} \right) + K_\frac{5}{2} \left( \frac{\pi l n}{NTR} \right) \right) \Bigg] \\
\xrightarrow{T\to 0} &- \frac{\pi^2 }{480 NR^4} - N^3T^4
\frac{\pi^2}{90} + T\frac{\pi}{2 R^3}e^{-\frac{\pi}{NTR}}.
\label{40}
\end{split}
\end{equation}
When $N=1$, this is one half of the  result  given in Eq.~(3.16)
in Ref.~\cite{schwinger78}, when dealing with the electromagnetic
field. The factor of 1/2 difference is what we would expect, since
we are considering the scalar field. The more general case with
refractive index $N$ was recently considered in
Ref.~\cite{brevik08}.

We evaluate the other part of the free energy (pertaining to the
extra dimensions) in the same way. Making use of
  Eq.~(\ref{29}) for the Epstein-like functions, we obtain the following final expression for the Casimir pressure (details omitted):
\begin{equation}
\begin{split}
P=& - \frac{\pi^2 }{480 NR^4} - \frac{1}{8 L^2 R \pi^2 N} \sum_{k=0}^{q-1} \binom{q}{k} (2)^{q-k} \\
 & \times \sum_{ m_1,m_2,\ldots,m_{q-k},n=1}^\infty
  \Bigg[ \frac{1}{R n^2} \left( \sum_{j=1}^{q-k} m_j^2 \right) K_2\left(2 n \frac{R}{L} \sqrt{ \sum_{j=1}^{q-k} m_j^2 } \right) \\
&~+ \frac{1}{L n} \left( \sum_{j=1}^{q-k} m_j^2 \right)^\frac{3}{2} \left( K_1\left(2 n \frac{R}{L} \sqrt{ \sum_{j=1}^{q-k} m_j^2 } \right) + K_3\left(2 n \frac{R}{L} \sqrt{ \sum_{j=1}^{q-k} m_j^2 } \right) \right) \Bigg]\\
&-N^3T^4 \frac{\pi^2}{90} + T\frac{\pi}{2 R^3}e^{-\frac{\pi}{NTR}}
- 2q T\left(\frac{NT}{2\pi L}\right)^{\frac{3}{2}}
e^{-\frac{1}{NTL}} \\
 & + q \frac{\pi T}{ R^3} \exp \left(
-\frac{1}{NT} \sqrt{ \frac{\pi^2}{R^2} + \frac{1}{L^2} }
\right). \label{42}
\end{split}
\end{equation}
When $N=1$, this expression agrees with Teo \cite{teo09} (she does
not restrict all  separations in the extra space to be equal). The
$T=0$ terms arising from the compactified dimensions are negative,
corresponding to an attractive force. The temperature corrections
(the last three terms) are small in the low temperature limit.

\subsection{High temperatures}

We shall consider only briefly the case of high temperatures,
\begin{equation}
RT \gg 1. \label{43}
\end{equation}
For instance, in order to satisfy this condition with one percent
accuracy when the  separation is $R= 1000$ nm, the temperature
must be quite high, $T> 200000$ K.  At such temperatures ordinary
solid bodies do not exist. This limit is thus of less physical
interest than that of low temperatures. Let us consider, however,
the surface force density. Our starting point is again
Eq.~(\ref{27}), from which we extract terms pairwise. Omitting all
details we present only the final result for the high-temperature
limit of the Casimir pressure:
\begin{equation}
\begin{split}
P =& -\frac{T \zeta_R(3)}{8 \pi R^3} - \frac{2\pi T^3 N^2}{R} e^{-4\pi RNT} -\frac{qT}{2\pi RL^2} e^{-2\frac{R}{L}} \\
&-\frac{qT}{\pi R L^2} \left( 1 + (2\pi NT L)^2 \right) \exp \left( -2 \frac{R}{L} \sqrt{ 1 + (2\pi NT L)^2} \right). \label{47}
\end{split}
\end{equation}
When $N=1$ our results are in accordance with Ref.~\cite{teo09}
though here we have included exponentially small corrections as
well, not just the terms linear in  temperature. In
Ref.~\cite{teo09} the size of each extra compacitifed dimension is
assumed arbitrary.

\section{Entropy, and the Nernst theorem}

Of  interest is  here the low temperature limit. To evaluate the
entropy $S=-\partial F/\partial T$ we extract once again terms
from the expression (\ref{27}) pairwise. As above we abstain from
giving the details of the calculation and present only the result:
\begin{equation}
\begin{split}
&S= -\frac{\partial}{\partial T} \Bigg\{ \sum_{k=0}^{q-1} \binom{q}{k} (2)^{q-k} \Bigg[ -\frac{T \sqrt{2\pi NT}}{2\pi^2} \sum_{\substack{m_1,m_2,\ldots,m_{q-k},\\n,l=1}}^\infty l^{-\frac{3}{2}} \left( \frac{\pi^2n^2}{R^2} +\sum_{j=0}^{q-k} \frac{m_j^2}{L^2} \right)^\frac{3}{4} \\
&\qquad \times K_\frac{3}{2} \left( \frac{l}{NT} \left( \frac{\pi^2n^2}{R^2} + \sum_{j=0}^{q-k} \frac{m_j^2}{L^2} \right)^\frac{1}{2} \right)\\
&\quad + \frac{RT^2N}{2\pi^2} \sum_{\substack{m_1,m_2,\ldots,m_p,\\n,l=1}}^\infty l^{-2} \left( \sum_{j=0}^{q-k} \frac{m_j^2}{L^2} \right)^\frac{1}{2} K_3 \left( \frac{l}{TL} \left(\sum_{j=0}^{q-k} m_j^2 \right)^\frac{1}{2} \right) \Bigg]\\
&\quad + N^3T^4 \frac{R \pi^2 }{90}-\frac{1}{\sqrt{2}} T \sqrt{
\frac{NT}{R^3} }
 \sum_{n,l=1}^\infty \left( \frac{n}{l} \right)^{\frac{3}{2}} K_{\frac{3}{2}} \left( \frac{\pi l n}{NTR} \right) \Bigg\}
\end{split} \label{48}
\end{equation}
(terms that do not contribute in $F$ are omitted). As the
derivatives of $K_\nu$ will produce terms containing $K_\nu$ with
the same argument we see that, when $T\rightarrow 0$, the $K_\nu$
will decay exponentially implying that $S\rightarrow 0$. The
Nernst theorem is satisfied. This is as we should expect, in the
present case with idealized boundary conditions. The same property
is known to hold for a metal without extra dimensions, when ideal
boundary conditions are assumed for all frequencies. (The current
discussion about thermal Casimir corrections relates to the case
where {\it dissipation} is included; for a discussion on these
issues see, for instance, Ref.~\cite{brevik06}.)

\section{On dispersive properties, when $p=q=1$}

As the second novel element in our analysis we will give a
simplified analysis of the change in free energy due to frequency
dispersion in the medium filling the physical space. As before, we
assume $p=1$ so that there are two infinite plates in physical
space separated by a gap $R$. To simplify the formalism we assume
moreover that $q=1$, so that there is only one extra dimension
(containing a vacuum, not a medium). This formal
simplification will still allow us to show the characteristics of
the dispersive behavior.

Now write the free energy $F$ as a sum of its $q=0$ part (i.e.,
ordinary space, with no extra dimension) and a $q=1$ part,
\begin{equation}
F=F_{q=0}+F_{q=1}. \label{26A}
\end{equation}
Here the first term is given by Eq.~(\ref{24}), for arbitrary
temperature. The Matsubara frequencies are $\zeta_l=2\pi lT$ with
$l=0,1,2,...$ The first finite value, corresponding to $l=1$, is
in dimensional terms $\zeta_1=2\pi k_BT/\hbar=2.4\times
10^{14}~$s$^{-1}$. This value is quite high, and it is therefore
physically reasonable, as mentioned in the Introduction, to ignore
the dispersive properties for higher values of $l$. We accordingly
adopt in the following a  dispersive model  in which
$N(\zeta)=N_0$ = constant for $l=1$ in the Epstein-like function
$E_2$, and then falls off abruptly such that we may put $N(\zeta)=1$ for $l \geq 2$. This point simplifies the
formalism, but still allows us to see the characteristics of the
dispersive behavior.

Starting from the general expression for $E_2$ we can thus make
the following effective substitution
\[ E_2\left((2\pi
NT)^2,\frac{\pi^2}{R^2};-1\right)=\sum_{n,l=1}^\infty
 \left[(2\pi NT)^2l^2+\frac{\pi^2}{R^2}n^2\right]
\]
\begin{equation}
\rightarrow E_2\left( (2\pi
T)^2,\frac{\pi^2}{R^2};-1\right) + E_1^{2\pi T N_0}\left(\frac{\pi^2}{R^2};-1\right)
- E_1^{2\pi T}\left(\frac{\pi^2}{R^2};-1\right), \label{27A}
\end{equation}
where
\begin{equation}
E_V^c(a_1,a_2,\ldots ,a_V;s)= \sum_{\{n_j\}=1}^\infty \left( c^2 + \sum_{j=1}^V a_j n_j^2 \right)^{-s} \label{28A}
\end{equation}
is a  generalized Epstein-like zeta function.
The first term in Eq.~(\ref{24}) ($F_{q=0}$) does not
refer to the dielectric properties of the physical space.
As for the first term and all other terms not containing $N(\zeta)$
we get no change from the previous sections.

Then, for the third term with $E_1$ in Eq.~(\ref{24}) we make use
of the same effective substitution as above, extracting the $l=1$
term,
\begin{equation}
E_1\left( (2\pi NT)^2; -\frac{3}{2}\right) \rightarrow E_1\left((2\pi T)^2;
-\frac{3}{2}\right) +(2\pi T)^3(N_0^3-1) \label{29A}
\end{equation}
No regularization of the $N_0$ term is here necessary.
Combining all three terms in $F_{q=0}$ we get
\begin{equation}
\begin{split}
F_{q=0} =& F_{q=0}^{N=1}
-\frac{T}{4 \pi} \left( E_1^{2\pi T N_0}\left(\frac{\pi^2}{R^2};-1\right)
- E_1^{2\pi T}\left(\frac{\pi^2}{R^2};-1\right) \right) \\
&+ \frac{TR}{6\pi} (2\pi T)^3(N_0^3-1).
\end{split}
\label{30A}
\end{equation}
Here we have collected all terms that give the free energy without
dispersion, i.e. $F_{q=0}^{N=1}$ is equal to Eq.~(\ref{24}) with
$N=1$ for all $l$. Hence, we can use all expressions obtained
above and only concentrate here on the additional terms introduced
by the  dispersion model. Since $E_1^{2\pi T N_0}(\pi^2/R^2;-1) =
\pi^2/R^2 \zeta_{EH}((2RTN_0)^2;-1) $ where $\zeta_{EH}$ is the
Epstein-Hurwitz zeta function, we might recall the conventional
expansion
\[ \zeta_{EH}(p;s)=\sum_{n=1}^\infty (n^2+p)^{-s} \]
\[
=-\frac{1}{2}p^{-s}+\frac{\sqrt{\pi}\Gamma(s-1/2)}{2\Gamma(s)}\,p^{-s+1/2}
\]
\begin{equation}
+\frac{2\pi^sp^{-s/2+1/4}}{\Gamma(s)}\sum_{n=1}^\infty
n^{s-1/2}\,K_{s-1/2}(2\pi n\sqrt{p}); \label{31A}
\end{equation}
cf.  \cite{elizalde95} (p. 81). Although this series is for many
purposes a convenient expression, it is in our case with small
arguments  better to write $\zeta_{EH}$ in another way: as $-s$ is
an integer we make use of Eq.~(3.24) on p.~37 in
Ref.~\cite{elizalde94} \footnote{Equation~(3.24) in
\cite{elizalde94} p. 37 gives three different expressions for the
Epstein-Hurwitz formula. One is for $-s \notin \mathbb{N}$ and
$1/2-s \notin \mathbb{N}$, the second for $-s \in \mathbb{N}$, and
the third for $1/2-s \in \mathbb{N}$. According to our
recalculations two of these expressions are correct, but  in the
case ($-s \in \mathbb{N}$) the expression should be different and
as given above. See the Appendix. It should also be noted that an
assumption for these expressions is that $c^2$ is small. This
assumption is fulfilled in the low temperature limit since $c^2 =
(2 RT N_0)^2 \ll 1$.} to write
\begin{equation}
\begin{split}
&E_1^{c}(1;s) = \zeta_{EH}(c^2;s) \\
&= \sum_{k=0}^{-s-1} (-1)^k \frac{\Gamma(k+s)}{k! \Gamma(s)} \zeta_R(2k+2s) c^{2k}
-\frac{\pi^{1/2}}{2\Gamma(s)} \Gamma \left( s-\frac{1}{2} \right) c^{1-2s}.
\end{split}
\label{32A}
\end{equation}
With this we find the finite expression
\begin{equation}
\begin{split}
&-\frac{T\pi}{4 R^2} \Gamma(-1)\zeta_{EH}((2RTN_0)^2;-1) \\
=& -\frac{T\pi}{4 R^2} \left( \Gamma(-1)\zeta_R(-2) - \frac{\pi^{1/2}}{2} \Gamma \left( -\frac{3}{2} \right) (2RTN_0)^3 \right) \\
= &-\frac{T\zeta_R(3)}{8\pi R^2} + \frac{4\pi^2}{3}RT^4 N_0^3.
\end{split}
\label{33A}
\end{equation}
We can now insert these expressions into Eq.~(\ref{30A}) to
evaluate $F_{q=0}$ at arbitrary temperature. Moreover, to simplify the
formalism somewhat, we restrict ourselves to the low temperature
approximation, $RT \ll 1$. From Eq.~(\ref{33}), (\ref{30A}) and (\ref{33A}) we find
\begin{equation}
F_{q=0}=\frac{8\pi^2}{3}RT^4(N_0^3-1)-\frac{\pi^2}{1440R^3}+\frac{\pi^2RT^4}{90}
-T^2 \frac{e^{-\frac{\pi}{RT}} }{2R} ,
\quad RT \ll 1, \label{34A}
\end{equation}
where the influence from the medium is shown explicitly in the
first term.

Consider next the term coming from the extra dimension, $q=1$.
From Eq.~(\ref{27}) we first get
\[ F_{q=1}=\frac{RT}{6\pi}E_1\left(\frac{1}{L^2};
-\frac{3}{2}\right)-\frac{T}{4\pi}\Gamma(-1)E_2\left(\frac{\pi^2}{R^2},\frac{1}{L^2};-1\right)
\]
\begin{equation}
+\frac{RT}{3\pi}E_2\left((2\pi NT)^2,
\frac{1}{L^2};-\frac{3}{2}\right)-\frac{T}{2\pi}\Gamma(-1)E_3\left((2\pi
NT)^2,\frac{\pi^2}{R^2},\frac{1}{L^2};-1\right), \label{35A}
\end{equation}
where the terms are written in order of increasing complexity. The
two last terms depend on $N$ and can be rewritten in the same manner as
Eq.~(\ref{27A}). Again, we write the free energy as the free energy
of a medium with $N=1$ (for all $l$) plus correction terms from $l=1$,
\begin{equation}
\begin{split}
F_{q=1} =& F_{q=1}^{N=1}
+ \frac{RT}{3 \pi} \left( E_1^{2\pi T N_0}\left(\frac{1}{L^2};-\frac{3}{2} \right)
- E_1^{2\pi T}\left(\frac{1}{L^2};-\frac{3}{2} \right) \right)\\
&-\frac{T}{2 \pi} \Gamma(-1) \left( E_2^{2\pi TN_0}\left(\frac{\pi^2}{R^2},\frac{1}{L^2};-1\right)  -E_2^{2\pi T}\left(\frac{\pi^2}{R^2},\frac{1}{L^2};-1\right) \right).
\end{split}
\label{36A}
\end{equation}
From p. 48 in \cite{elizalde95} we adopt a formula similar to
Eq.~(\ref{29}) (with some re-writing)
\begin{equation}
\begin{split}
&E_V^c(a_1,a_2,\ldots,a_V;s) = -\frac{1}{2} E_{V-1}^c(a_2,a_3,\ldots ,a_V;s) \\
&\quad + \frac{1}{2} \sqrt{\frac{\pi}{a_1}} \frac{\Gamma \left(s-\frac{1}{2} \right)}{\Gamma (s)} E^c_{V-1}(a_2,a_3,\ldots,a_V;s-\frac{1}{2}) \\
& \quad + \frac{2 \pi^s}{\Gamma(s)} a_1^{-(s+1/2)/2} \sum_{n_1,n_2,\ldots,n_V=1}^\infty n_1^{s-1/2} \left( c^2+ \sum_{i=2}^V a_i n_i^2 \right)^{-(s-1/2)/2} \\
& \qquad \times K_{s-1/2}\left( \frac{2\pi}{\sqrt{a_1}} n_1 \left( c^2 + \sum_{i=2}^V a_i n_i^2 \right)^{1/2} \right).
\label{37A}
\end{split}
\end{equation}
Then, we can write
\begin{equation}
\begin{split}
&\Gamma(-1)E_2^{2\pi TN_0}\left(\frac{\pi^2}{R^2},\frac{1}{L^2};-1\right)\\
&=-\frac{1}{2}\Gamma(-1)E_1^{2\pi TN_0}\left(\frac{1}{L^2};-1\right)
+\frac{R}{2\sqrt{\pi}} \Gamma \left(-\frac{3}{2} \right) E_1^{2\pi TN_0}\left(\frac{1}{L^2};-\frac{3}{2}\right)\\
&+\frac{2}{\sqrt{\pi R}}\sum_{m,n=1}^{\infty} n^{-3/2} \left( \frac{m^2}{L^2}+ (2\pi TN_0)^2 \right)^{3/4} K_{3/2}\left( 2R n \sqrt{\frac{m^2}{L^2}+ (2\pi TN_0)^2} \right).
\end{split}
\label{38A}
\end{equation}
There occurs a cancellation since
$-T/(2\pi)(R/(2\sqrt{\pi}))\Gamma(-3/2)E_1^{2\pi TN_0}(1/L^2;-3/2)= -RT/(3\pi)E_1^{2\pi TN_0}(1/L^2;-3/2)$. With the aid of Eq.~(\ref{32A}) and the reflection
formula for $\zeta_R$ we see that
\begin{equation}
-\frac{1}{2}\Gamma(-1)E_1^{2\pi TN_0}\left( \frac{1}{L^2},-1\right)
=-\frac{\zeta_{R}(3)}{4\pi^2 L^2}+\frac{8}{3}\pi^4 T^3 L N_0^3.
\label{39A}
\end{equation}
When all is put together, we get
\begin{equation}
\begin{split}
&F_{q=1}=F_{q=1}^{N=1} - \frac{4}{3}\pi^3 T^4 L (N_0^3-1)\\
&-\frac{T}{\sqrt{\pi^3 R}} \sum_{m,n=1}^{\infty} n^{-3/2} \left( \frac{m^2}{L^2}+ (2\pi TN_0)^2 \right)^{3/4} K_{3/2}\left( 2R n \sqrt{\frac{m^2}{L^2}+ (2\pi TN_0)^2} \right) \\
&+\frac{T}{\sqrt{\pi^3 R}} \sum_{m,n=1}^{\infty} n^{-3/2} \left( \frac{m^2}{L^2}+ (2\pi T)^2 \right)^{3/4} K_{3/2}\left( 2R n \sqrt{\frac{m^2}{L^2}+ (2\pi T)^2} \right).
\end{split}
\label{40A}
\end{equation}
Since $R/L \gg 1$ we only need to take into account the $n=m=1$ term of the double-sum.
Some calculation yields for the medium-independent $F_{q=1}^{N=1}$-terms
\[ \Gamma(-1)E_3\left((2\pi
T)^2,\frac{\pi^2}{R^2},\frac{1}{L^2};-1\right)=-\frac{1}{2}\Gamma(-1)E_2\left(
 \frac{\pi^2}{R^2},\frac{1}{L^2};-1\right) \]
\[-\frac{1}{720TL^3}+ \frac{3R\zeta_R(5)}{32 \pi^5 T L^4}
+ \frac{1}{2\pi RTL^2} \sum_{m,n=1}^\infty \frac{m^2}{n^2} K_2 \left( \frac{2nmR}{L} \right) \]
\begin{equation}
+\frac{2\sqrt{2\pi T}}{\pi}\sum_{l,n,m=1}^\infty
l^{-3/2}\left(\frac{\pi^2 n^2}{R^2}+\frac{m^2}{L^2}\right)^{3/4}K_{3/2}\left(
\frac{l}{T}\sqrt{ \frac{\pi^2 n^2}{R^2}+\frac{m^2}{L^2} }\right).
\label{41A}
\end{equation}
and
\[ E_2\left( (2\pi
T)^2,\frac{1}{L^2};-\frac{3}{2}\right)=-\frac{1}{240L^3}+\frac{9}{64
\pi^5 TL^4}\zeta_R(5) \]
\begin{equation}
+\frac{3T}{\pi L^2}\sum_{l,m=1}^\infty \frac{m^2}{l^2}K_2\left(
\frac{lm}{LT}\right). \label{42A}
\end{equation}
At last we have the final low temperature expression for $L/R \ll 1$
\begin{equation}
\begin{split}
F_{q=1}=&-\frac{4\pi^3}{3}LT^4(N_0^3-1)+\frac{1}{1440\pi
L^3}-\frac{1}{8(\pi RL)^{3/2}} e^{-2R/L}\\
&-\frac{T^2}{\pi}\sqrt{\frac{\pi^2}{R^2} + \frac{1}{L^2}} e^{-\frac{1}{T}\sqrt{\frac{\pi^2}{R^2} + \frac{1}{L^2}} }\\
&-\frac{T}{2\pi R}\sqrt{\frac{1}{L^2}+(2\pi TN_0)^2} e^{-2R\sqrt{\frac{1}{L^2}+(2\pi TN_0)^2}}\\
&+\frac{T}{2\pi R}\sqrt{\frac{1}{L^2}+(2\pi T)^2} e^{-2R\sqrt{\frac{1}{L^2}+(2\pi T)^2}}
, \quad RT \ll 1, ~ L/R \ll 1.
\end{split}
\label{43A}
\end{equation}
Adding Eq.~(\ref{34A}) and (\ref{43A}) we get
the total free energy per unit surface in physical space to the
lowest order in $RT$ and $R/L$
\begin{equation}
\begin{split}
F=&\frac{1}{1440\pi L^3}\left[ 1-\left(\frac{\pi
L}{R}\right)^3\right] +\frac{\pi^2}{90}RT^4 +(\frac{2R}{\pi}-L)\frac{4}{3}\pi^3 T^4 (N_0^3-1)\\
&-T^2 \frac{e^{-\frac{\pi}{RT}} }{2R}-\frac{1}{8(\pi RL)^{3/2}} e^{-2R/L} -\frac{T^2}{\pi}\sqrt{\frac{\pi^2}{R^2} +
\frac{1}{L^2}} e^{-\frac{1}{T}\sqrt{\frac{\pi^2}{R^2} + \frac{1}{L^2}} }\\
&-\frac{T}{2\pi R}\sqrt{\frac{1}{L^2}+(2\pi TN_0)^2} e^{-2R\sqrt{\frac{1}{L^2}+(2\pi TN_0)^2}}\\
&+\frac{T}{2\pi R}\sqrt{\frac{1}{L^2}+(2\pi T)^2} e^{-2R\sqrt{\frac{1}{L^2}+(2\pi T)^2}}
, \quad RT \ll 1, ~ L/R \ll 1.
\end{split}
\label{44A}
\end{equation}
For the pressure $P=-\partial F/\partial R$ we obtain, when
omitting the small correction terms,
\begin{equation}
\begin{split}
P=&-\frac{\pi^2}{480R^4} - \frac{\pi^2 T^4}{90} \left[ 240(N_0^3-1) + 1 \right]\\
&+T \frac{e^{-\frac{\pi}{RT}} }{2R^3} -\frac{1}{4L(\pi RL)^{3/2}} e^{-2R/L}
+\frac{\pi T}{R^3} e^{-\frac{1}{T} \sqrt{\frac{\pi^2}{R^2}+\frac{1}{L^2}} } \\
&-\frac{T}{\pi R}\left(\frac{1}{L^2}+(2\pi TN_0)^2\right) e^{-2R\sqrt{\frac{1}{L^2}+(2\pi TN_0)^2}} \\
&+\frac{T}{\pi R}\left(\frac{1}{L^2}+(2\pi T)^2\right) e^{-2R\sqrt{\frac{1}{L^2}+(2\pi T)^2}}.
\end{split}
\label{45A}
\end{equation}
Here the first term gives the conventional attractive force in
physical space. The second term is attractive and can be traced
back to the $q=0$ contribution. In the force expression there are
Bessel functions with the same arguments as in the free energy.
There are several sums over Bessel functions with different
arguments; we have kept the largest term from each sum to show the
relative sign and size. Note that since $N_0 > 1$ the contribution
from the extra dimension is attractive. It is seen that cross
terms, containing contributions from both the $R$ and the $L$
spaces, are all exponentially small.

\section{Summary}

We began by developing the  finite temperature Casimir theory of
the scalar field, assuming Dirichlet boundary conditions for
 two large parallel plates with a gap $R$ in physical
space endowed with a medium with constant 'refractive index' $N$,
and in addition an arbitrary number $q$ of extra compactified
dimensions.  Zeta function regularization of the Epstein-like
functions was used throughout. In the main part of the formalism
the radii $L$ of the compactified dimensions were taken to be
equal. For general values of $T$, the free energy $F$ was given by
Eq.~(\ref{27}). For low $T$ (i.e., $RT \ll 1$) which was the case
of main physical interest, the approximate expression for $F$ was
given by Eq.~(\ref{36}), and the corresponding Casimir pressure
was given by Eq.~(\ref{42}). When $N=1$, agreement was found with
Teo \cite{teo09}. The Nernst theorem was found to be satisfied;
cf.   the low temperature entropy expression (\ref{48}). The case
of high temperatures was briefly discussed. In all cases of
physical interest, the inequality $R/L \gg 1$ could be assumed.

In the dispersive part of the theory considered in Section~6
(assuming $q=1$), we introduced a rough model implying neglect of
the dispersive property at all frequencies higher than the first
Matsubara frequency, $\zeta_1=2\pi T$. It is worth noticing that for the Epstein-Hurwitz zeta function
 we found it convenient to use the expansion (\ref{32A}), instead of the more conventional expansion
 (\ref{31A}). We found (\ref{32A}) to be the most suitable form for low temperatures.
For  $RT \ll 1$ we calculated the free energy  in Eq.~(\ref{44A}),
showing contributions from physical space as well as from the
extra dimension. The corresponding pressure in Eq.~(\ref{45A})
contains both attractive and repulsive terms. The common property
of the cross terms, i.e.  terms containing both $R$ and $L$
contributions, is that these terms are exponentially small.

\bigskip

{\bf Acknowledgment}

\bigskip

We thank Emilio Elizalde for information regarding the issue discussed in the Appendix.

\newpage
\renewcommand{\theequation}{\mbox{\Alph{section}.\arabic{equation}}}
\appendix
\setcounter{equation}{0}

\section{Regularization of the Epstein-Hurwitz zeta function}
In this appendix we will argue why  the correct
regularized expression for the Epstein-Hurwitz zeta function should be
\begin{equation}
\begin{split}
&E_1^{c}(1;s) = \zeta_{EH}(c^2;s) \\
&= \sum_{k=0}^{-s-1} (-1)^k \frac{\Gamma(k+s)}{k! \Gamma(s)} \zeta_R(2k+2s) c^{2k}
-\frac{\pi^{1/2}}{2\Gamma(s)} \Gamma \left( s-\frac{1}{2} \right) c^{1-2s}
\end{split}
\label{app1}
\end{equation}
for $-s \in \mathbb{N}$. As noted above, this expression is a bit different
from the one given in Eq.~(3.24) in \cite{elizalde94}:
\begin{equation}
\begin{split}
&E_1^{c}(1;s) =   \sum_{k=0}^{-s} (-1)^k \frac{\Gamma(k+s)}{k!
\Gamma(s)} \zeta_R(2k+2s) c^{2k}.
\end{split}
\label{app2}
\end{equation}
It is seen that the two expressions differ with respect to two
terms: the first is $k=-s$ from the sum over $k$, the second is a
term proportional to $c^{1-2s}$. Referring to the basic
calculation in \cite{elizalde94} we will argue why the $k=-s$ term
should be omitted and the $c^{1-2s}$ term should be present.

We start with (3.7) in \cite{elizalde94}
\begin{equation}
E_1^c(1;s) = \sum_{\substack{k=0 \\ k \neq 1/2-s ~\mathrm{if}~1/2-s \in \mathbb{N} \\ k \neq -s ~\mathrm{if}~-s \in \mathbb{N} }}^\infty
(-1)^k \frac{\Gamma(k+s)}{k! \Gamma(s)} \zeta_R(2k+2s) c^{2k}+ \mathrm{Res}_z,
\label{app3}
\end{equation}
where
\begin{equation}
\mathrm{Res}_z  \equiv \mathrm{Res} \left[ \phi(a)=\frac{\Gamma(s+a)}{\Gamma(a+1)\Gamma(s)} c^{2a} \zeta_R(2s+2a)\pi \csc\pi a; a=z\right]
\label{app4}
\end{equation}
for $z=1/2-s$ and $z=-s$. (The function $\phi (a)$ comes from the
asymptotic behaviour of the integrand when the sum is converted
into an integral; cf. \cite{elizalde94}.) The sum over $k$ runs
from 0 to $\infty$, but skips $k=1/2-s$ (if $1/2-s \in
\mathbb{N}$) to avoid the pole $\zeta_R(1)$ and $k=-s$ (if $-s \in
\mathbb{N}$) to avoid the pole $\Gamma(0)$. The next step is to
calculate the residue at $a=1/2-s$ and $a=-s$. We  quote the
results from \cite{elizalde94}:
\begin{equation}
\begin{split}
&{\rm Res}_{1/2-s} \\
&= \left\{
\begin{array}{ll}
  -\frac{\pi^{1/2}}{2\Gamma(s)}\Gamma \left(s-\frac{1}{2}\right) c^{1-2s} & \text{if } 1/2-s\notin \mathbb{N},\\
  \begin{split}
  &\frac{(-1)^{-(s-1/2)} \pi^{1/2}}{\Gamma(s) \Gamma \left(\frac{3}{2}-s\right)} c^{1-2s}  \\
  &\times \left[\frac{1}{2}\left( \psi  \left(\frac{1}{2}\right) - \psi  \left(\frac{3}{2}-s\right)
  + \log c^2 \right) + \gamma \right]
  \end{split}
  & \text{if } 1/2-s \in \mathbb{N},
\end{array} \right.
\end{split}
\label{app5}
\end{equation}
where $\psi(a) \equiv \frac{\mathrm{d}}{\mathrm{d}a} \log \Gamma(a)$.
For $a=-s$ we have
\begin{equation}
\mathrm{Res}_{-s} =
\left\{
\begin{array}{ll}
\frac{1}{2}c^{-2s} & \text{if } -s \notin \mathbb{N}\\
0 & \text{if } -s \in \mathbb{N}.
\end{array}
\right.
\label{app6}
\end{equation}
The results can be divided into three: $1/2-s \notin \mathbb{N}$
and $-s \notin \mathbb{N}$; $1/2-s \in \mathbb{N}$; and $-s \in
\mathbb{N}$. We are interested in $-s \in \mathbb{N}$. Combining
Eqs.~(\ref{app3}), (\ref{app5}) and (\ref{app6}) we find this to
be equal to Eq.~(\ref{app1}). The sum over $k$ only runs to $-s-1$
since all terms with $k >-s$ are equal to zero since
$\zeta_R(-2n)=0$ for $n=1,2,3\ldots$ and $k=-s$ is  excluded for
$-s \in \mathbb{N}$. The $c^{1-s}$ term must be included since $-s
\in \mathbb{N}$ implies $1/2-s \notin \mathbb{N}$.

\end{document}